\documentclass[twocolumn, aps, prd, 10pt, superscriptaddress, nofootinbib]{revtex4-2}

\usepackage{microtype}
\usepackage[T1]{fontenc}
\usepackage{amsmath}
\usepackage{fontawesome}
\usepackage{graphicx}
\usepackage[dvipsnames]{xcolor}
\usepackage{hyperref}
  \hypersetup{  
    colorlinks=true,
    linkcolor=BrickRed,
    filecolor=BrickRed,
    citecolor=BrickRed,
    urlcolor=RoyalBlue,
  }
\usepackage[figure*]{hypcap}
\usepackage{aas_macros}

\newcommand{\dee}{\mathrm{d}}
\renewcommand{\vec}[1]{\boldsymbol{#1}}

\begin{document}


\title{Field-level Emulation of Cosmic Structure Formation\\with Cosmology and Redshift Dependence}

\author{Drew Jamieson}
\email{jamieson@mpa-garching.mpg.de}
\affiliation{Max-Planck-Institut f\"ur Astrophysik, Karl-Schwarzschild-Straße 1, 85748 Garching, Germany}

\author{Yin Li}
\affiliation{Department of Mathematics and Theory, Peng Cheng Laboratory, Shenzhen, Guangdong 518066, China}

\author{Francisco Villaescusa-Navarro}
\affiliation{Center for Computational Astrophysics, Flatiron Institute,
162 5th Avenue, New York, NY 10010, USA}

\author{Shirley Ho}
\affiliation{Center for Computational Astrophysics, Flatiron Institute,
162 5th Avenue, New York, NY 10010, USA}

\author{David N. Spergel}
\affiliation{Center for Computational Astrophysics, Flatiron Institute,
162 5th Avenue, New York, NY 10010, USA}
\affiliation{Simons Foundation, 160 5th Avenue, New York, NY 10010, USA}

\date{\today}

\begin{abstract}
We present a field-level emulator for large-scale structure, capturing the cosmology dependence and the time evolution of cosmic structure formation. The emulator maps linear displacement fields to their corresponding nonlinear displacements from N-body simulations at specific redshifts. Designed as a neural network, the emulator incorporates style parameters that encode dependencies on $\Omega_{\rm m}$ and the linear growth factor $D(z)$ at redshift $z$. We train our model on the six-dimensional N-body phase space, predicting particle velocities as the time derivative of the model's displacement outputs. This innovation results in significant improvements in training efficiency and model accuracy. Tested on diverse cosmologies and redshifts not seen during training, the emulator achieves percent-level accuracy on scales of $k\sim~1~{\rm Mpc}^{-1}~h$ at $z=0$, with improved performance at higher redshifts. We compare predicted structure formation histories with N-body simulations via merger trees, finding consistent merger event sequences and statistical properties.
\end{abstract}


\maketitle

\section{Introduction}
\label{sec:intro}

As cosmological data analysis pushes to smaller scales, utilizes higher-order statistics \cite{gualdi_22,Philcox:2021hbm,Hou:2022wfj,Philcox:2022hkh}, and implements field-level analyses \cite{SimBIG:2023ywd,Hahn:2022wgo,Stopyra:2023yqm,Andrews:2022nvv,Stadler:2023hea,Nguyen:2024yth,Babic:2024wph} and simulations-based inference schemes \cite{Alsing:2019xrx,Cranmer:2019eaq,Massara:2024cvu,Modi:2023llw,Tucci:2023bag,Lin:2022ayr}, there is a growing demand for computationally efficient and accurate methods of predicting the outcomes of nonlinear cosmic structure formation. Traditional analysis methods of N-point statistics beyond the power spectrum and bispectrum require large mock datasets for accurate covariance estimation. The same is true for alternative summary statistics such as k-nearest neighbours \cite{Yuan:2023ugr,Coulton:2023ouk,Wang:2021kbq} or scattering wavelets \cite{Cheng:2020qbx,Valogiannis:2021chp,Eickenberg:2022qvy,Valogiannis:2022xwu,SimBIG:2023gke}. Simulations-based inference methods and field-level analyses require generating many accurate realizations of the late-time density field for constraining model parameters and initial conditions reconstruction. As next-generation galaxy survey data become available from DESI \citep{{DESI:2016fyo,DESI:2024mwx}}, Euclid \citep{Amendola:2016saw}, LSST at the Vera C. Rubin Observatory \citep{LSST:2008ijt}, SPHEREx \citep{SPHEREx:2014bgr}, and Subaru Prime Focus Spectrograph \citep{PFSTeam:2012fqu}, obtaining the best possible constraints on cosmological parameters and initial conditions will require accelerated, highly accurate predictions for survey observables.

Recent advancements in machine-learning methods for modelling nonlinear cosmological structure formation have considerably improved computational efficiency and small-scale accuracy. These include error corrections for fast particle-mesh-based simulations \citep{Kaushal:2021hqv}, interpolation between two simulation snapshots \citep{chen2020learning}, generative methods for super-resolution simulations \citep{Doogesh_2020,Li:2020vor,Ni:2021mzk,schaurecker2021super} along with their redshift dependence \citep{Zhang:2023lqi}, normalizing flows \citep{Dai:2022dso,Dai:2023lcb}, and mapping between standard $\Lambda$CDM and modified gravity cosmologies \citep{Saadeh:2024vuj}. The first neural network emulating the N-body mapping at the field level, including cosmology dependence, was demonstrated in \citet{Jamieson:2022lqc}. This built upon previous work emulating particle-mesh simulations \citep{He:2018ggn,RenanEtAl2020}. The N-body neural network model was tested on a set of initial conditions, such as spherical collapse and pairs of isolated coupled modes, demonstrating its ability to accurately predict the physics of cosmic structure formation far outside of its training data \citep{Jamieson:2022daw}. This field-level emulator has been implemented in hybrid Lagrangian bias modelling of the galaxy field \citep{Ibanez:2023vxb} and Hamiltonian Monte Carlo Bayesian inference of initial conditions \citep{Doeser:2023yzv}.

In this work, we extend the field-level N-body emulator by adding redshift dependence and training on simulation snapshots at multiple redshifts. Due to the time dependence and autodifferentiablility of this new model, we can efficiently obtain the N-body particle velocities as the time derivative of the output particle displacements. Since we can evaluate these velocities on the fly during training, we can define a loss function that depends on both the particle positions and velocities, training on the six-dimensional N-body phase space. Enforcing the physical constraint that velocities must equal the displacement time derivatives improves training efficiency and increases the model's accuracy, especially for the velocity field. Our model is an example of physics-informed neural networks (PINNs) \citep{2019JCoPh.378..686R,2022arXiv221108064H}. We impose a prior on the relationship between the time-dependence of the particle coordinates and the particle velocities explicitly through our model design and, as we will describe later, through our choice loss function.

The emulator is extremely fast, capable of predicting the nonlinear displacement and velocity fields for $128^3$ particles in half a second on a single GPU. It is parallelized and scales well with multi-GPU processing, capable of generating realizations for an arbitrarily large box size. The model achieves percent-level accuracy down to scales of $k\sim~1~{\rm Mpc}^{-1}~h$.

This paper is structured as follows. In Section \ref{sec:model} we present the model architecture. In Section \ref{sec:training} we describe the training data and loss function. Section \ref{sec:res} contains the results of our model tested on simulations that were not in its training set. We conclude in Section \ref{sec:conc}.

\section{Model}
\label{sec:model}

We describe N-body particles by their coordinates, $\vec{x}$, in a periodic simulation box. Each particle is associated with a site $\vec{q}$ on a regular cubic lattice, so its position at redshift $z$ is defined as,
\begin{align}
    \vec{x}(\vec{q}, z) = \vec{q} + \vec{\Psi}(\vec{q}, z) \, .
\end{align}
Here $\vec{\Psi}$ is the displacement field, and $\vec{q}$ are the Lagrangian coordinates of the particle. In the linear, Zeldovich approximation (ZA), the displacement field evolves as,
\begin{align}
    \vec{\Psi}_{\rm ZA}(\vec{q}, z) = \frac{D(z)}{D(z_i)}\vec{\Psi}(\vec{q}, z_i) \, ,
\end{align}
where $D(z)$ is the linear growth factor, and $z_i$ is a redshift chosen early enough that the displacement field is well-described by linear theory. As the nonlinearity of gravitational clustering becomes important at late times, this linear approximation becomes inaccurate and nonperturbative methods of simulating cosmic structure formation, such as N-body simulations, become necessary (see \citet{2022LRCA....8....1A} for a review).

The full phase space of the N-body system includes the particle velocities,
\begin{align}
    \label{eq:vel}
    \dot{\vec{\Psi}}(\vec{q}, z) = D(z) f(z) H(z) \frac{\dee \vec{\Psi}(\vec{q}, z)}{\dee D(z)} \, .
\end{align}
Here $f(z) = \dee \log D(z) / \dee \log(a)$ is the linear growth rate, $a$ is the scale factor, and $H(z)$ is the Hubble rate. The particle velocities are needed to model redshift-space distortions in galaxy surveys. We designed our field-level emulator to predict the nonlinear particle displacements and velocities at any redshift ranging from $z=3$--$0$ based on the ZA displacement field at the target redshift.

\begin{figure*}
    \centering
    \includegraphics[width=\linewidth]{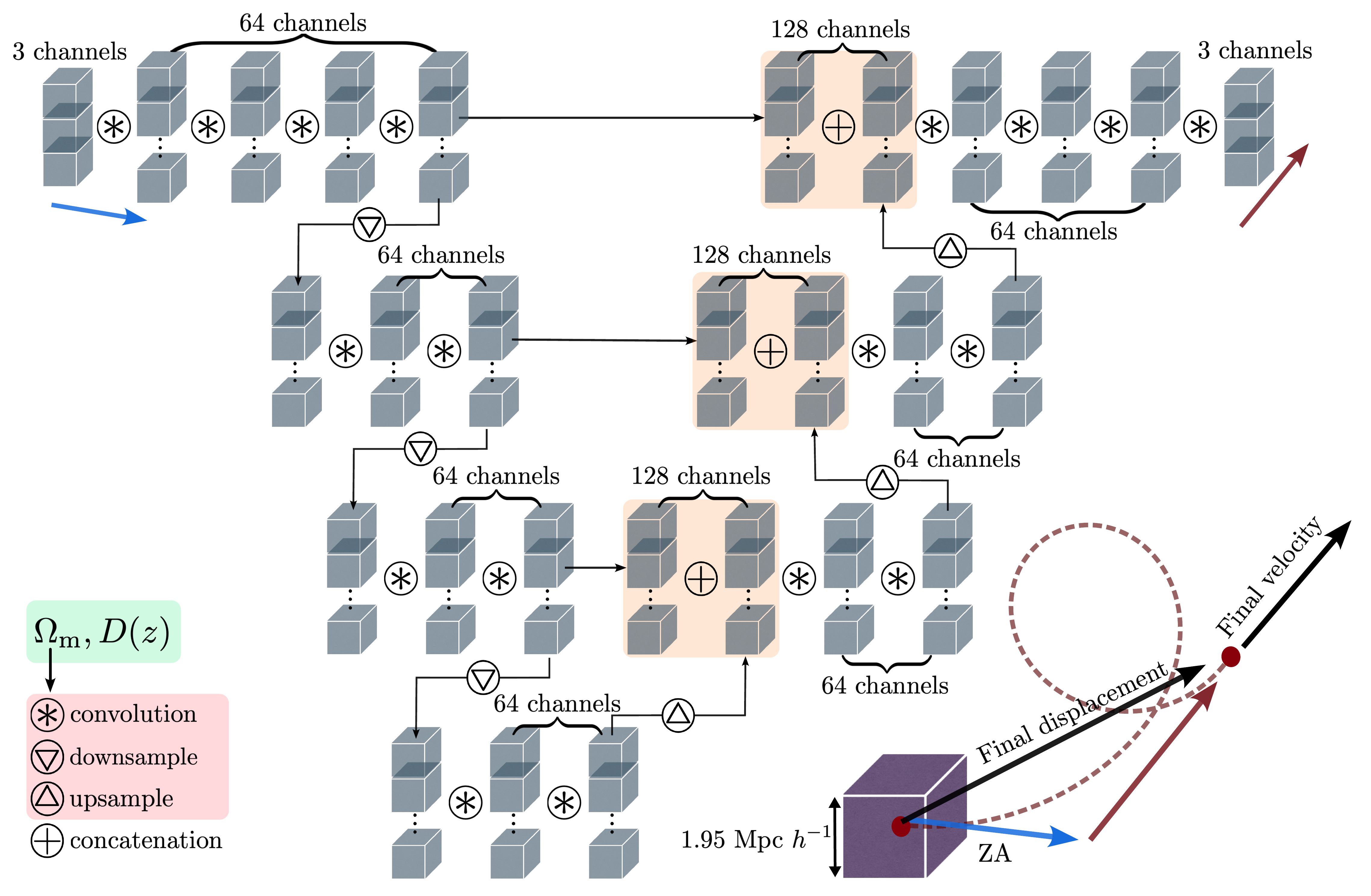}
    \caption{Here, we depict the architecture of the field-level emulator. In the bottom-right corner, we depict a single particle's Lagrangian cell. Its initial ZA displacement is the blue vector. The 3D vector field of ZA displacements is input for the model. The red dashed curve shows the particle's trajectory. The field-level emulator predicts the red vector, the residual between the linear and nonlinear displacements. The time derivative of the model's displacement output is the final velocity vector. All convolutional operations, including the downsampling and upsampling ones, are modulated by parameters that depend on $\Omega_{\rm m}$ and $D(z)$, allowing the emulator to capture cosmology and redshift dependence.}
    \label{fig:cnn}
\end{figure*}

The field-level emulator is a convolutional neural network augmented with style parameters. We implement and train the model in \texttt{PyTorch} \citep{pytorch}, using the \texttt{map2map} (\href{https://github.com/eelregit/map2map}{\faGithub} \url{github.com/eelregit/map2map}) code base. The model and its parameters can be found at \href{https://github.com/dsjamieson/m2m_nbody_zdep}{\faGithub} \url{github.com/dsjamieson/m2m_nbody_zdep}.

The neural network implements a U-Net/V-Net design, shown in Fig.~\ref{fig:cnn}. The input has three channels corresponding to the Cartesian components of the ZA displacements at the desired redshift, arranged in a 3D grid. The input passes through four ResNet 3$\times$3$\times$3 convolutions. The first of these convolutional operations transforms the 3 input channels to 64 internal channels. The subsequent convolutions map from 64 to 64 internal channels. After the four convolutional operations, a copy of the outcome is stored for use on the upsampling side of the network before the result is downsampled with a 2$\times$2$\times$2 convolutional kernel. The model repeats this sequence of operations two times, except these intermediate levels of the network use two instead of four convolutions before downsampling. After the fourth and deepest level of the network, we do not store a copy of the outcome. The model then upsamples the data with a 2$\times$2$\times$2 convolutional filter. The stored outcome from the previous level is then appended to the data, doubling the number of internal channels to 128 before convolving twice and upsampling again. The first of these convolutions maps the 128 channels back down to 64. The model repeats this once before the final level of the network, which performs four convolutions. The final operation maps the internal 64 channels to a three-channel final result. The model then adds the input to the output. This addition means that the output represents the Cartesian components of the difference between the nonlinear and linear displacements. See \citet{Jamieson:2022lqc} for additional details.

This architecture has a finite receptive field corresponding to 48 grid sites on either side of a given focal cell. We compute the receptive field's size based on the sequence of convolutions and downsampling/upsampling operations. Predicting a single particle's displacement\footnote{The downsampling and upsampling operations will not process data that is odd in dimension, so the smallest size box that the model can process is for $2^3$ particles, or $98^3$ with the padding.}, a box of size $97^3$, centred on the focal particle would need to be passed through the network, which corresponds to a Lagrangian volume of comoving size $(189.45~\mathrm{Mpc}~h^{-1})^3$. Since the network is missing information for all cells in the padding region, it cannot accurately predict the displacements for these particles and we must discard this padded region from the output. The network can process a box of even size as long as it is padded by $48$ cells on all sides. The dimensions of the input and output are $3 \times (96 + N)^3$, which we crop down to the desired size of $3 \times N^3$ by discarding the padding. Since the simulations in our training and testing datasets are run in periodic boxes, the padding must be done periodically. There is an advantage to the limited receptive field of the network: it preserves the ZA field on large scales where linear theory is accurate.

The network described so far can train on data from simulation snapshots at a single redshift for a fixed cosmology. To expand its capabilities, allowing the network to learn the N-body mapping as a function of $\Omega_{\mathrm{m}}$ and redshift, we augment it to include $\emph{style parameters}$ \citep{StyleGAN2}. Before any of the convolutions including downsampling/upsampling operations are performed, the values of $\Omega_{\mathrm{m}}$ and $D(z)$ for the snapshot are passed to the network and mapped to internal arrays matching the dimensions of the convolutional kernels. These parameters then modulate the network weights. In effect, this elevates the single neural network to a two-parameter family of neutral networks, continuously and autodifferentiably parameterized by the style inputs $\Omega_{\mathrm{m}}$ and $D(z)$. We use $D(z)$ to quantify time dependence because it makes the particle velocities easier to obtain from Eq.~\eqref{eq:vel}. The other $\Lambda$CDM cosmological parameters ($\sigma_8$, $n_s$, $\Omega_{\rm b}$, and $h$ ), affect only the input, through the shape of the matter transfer functions. The physics of gravitational clustering does not depend on these parameters, so they would not be informative to the neural network as it learns the N-body mapping and should not be included as input for the style parameters.

\section{Training}
\label{sec:training}

The style parameters and network parameters are trained simultaneously by randomly sampling snapshots from a suite of simulations with varied cosmological parameters and a fixed set of snapshot redshifts. We use the Quijote Latin hypercube simulations \citep{Villaescusa-Navarro:2019bje}, which were run with $512^3$ particles in a box of side length $1~\mathrm{Gpc}~h^{-1}$. The Lagrangian space resolution for all of these simulations is $1.95~\mathrm{Mpc}~h^{-1}$.  Since the emulator operates in Lagrangian space, it can only reliably make predictions for this specific resolution. There are 2000 simulations in the simulations dataset. Each simulation has a unique set of the five $\Lambda$CDM cosmological parameters $\Omega_{\mathrm{m}}\in [0.1,\ 0.5]$, $\Omega_{\mathrm{b}}\in[0.015,\ 0.035]$, $h\in[0.5,\ 0.9]$, $n_\mathrm{s}\in[0.85,\ 1.05]$, and $\sigma_8\in[0.6,\ 1.0]$ sampled on a 5D latin hypercube. Each simulation also has unique initial conditions. Snapshots are stored at redshifts $z\in\{0,\ 0.5,\ 1,\ 2,\ 3\}$.

We partition the 2000 simulations into three sets: 1874 for training, 122 for validation, and 4 for testing. The previous work \citep{Jamieson:2022lqc} had roughly equal numbers of testing and validation simulations. However, since the model errors depend only on the values of $\Omega_{\rm m}$ and $\sigma_8$, we take the most extreme values among these two parameters from the previous testing set and add one simulation from the Quijote fiducial cosmology (see \citep{Villaescusa-Navarro:2019bje}) for our testing data. In total, there are 5 test simulations. We added all other simulations to the training data. To encourage isotropy, we use data augmentation, randomly transforming the input and target data by the symmetries of a cube (see \citet{Jamieson:2022lqc} for details).

The most challenging part of the mapping to learn is when the nonlinearity is strongest, at redshift $z=0$. We already have a pretrained model for this redshift, so it is most efficient to initialize the new model to be nearly identical to the old model and have it learn the comparatively easier mapping at earlier redshifts. We accomplish this by initializing all network parameters with those from the pretrained, fixed-redshift model. Then we initialize the new style parameters randomly. We first trained only the style parameters for 10 epochs, holding the network parameters fixed. We then train all parameters for another 10 epochs.

Including redshift dependence has several benefits. Since the model is automatically differentiable with respect to its style inputs, the particle velocities can be obtained as the time derivatives of their displacements. Taking the forward mode derivative, or Jacobian-vector-product (JVP) \citep{GriewankWalther2008}, with respect to the style input $D(z)$, we use Eq.~\eqref{eq:vel} to obtain the model prediction for the velocities. The fixed-redshift model \citep{Jamieson:2022lqc} needed two separate neural networks: one for displacements and one for velocities. Now, we only need the displacement model, which eliminates the training time for the second model. This is already a significant advantage, but there is another benefit from adding redshift dependence. We can evaluate the velocities along with the displacements during training. This allows us to include terms involving the N-body particle velocities in the loss function. The model is then trained on the full N-body phase space. The network learns the trajectories of the particles as parametric curves depending on $D(z)$ and simultaneously learns that the velocities must be tangent vectors to the trajectories. This imposes a stringent physical constraint, requiring the particle velocities to equal the displacement time derivatives, which greatly improves training efficiency and model accuracy. This can also be viewed as multi-task learning \cite{zhang2021survey}.

The loss function we use contains four terms. The first is the mean squared error (MSE) for the particle displacements,
\begin{align}
    L_{\vec{\Psi}} = \frac{1}{N}\sum_{i=1}^{N} |\vec{\Psi}_{i,\mathrm{SNN}} - \vec{\Psi}_{i,\mathrm{SIM}}|^2 \, ,
\end{align}
where $\vec{\Psi}_{i,\mathrm{SNN}}$ is the emulator displacement prediction for the $i^{\rm th}$ particle and $\vec{\Psi}_{i,\mathrm{SIM}}$ is the true N-body displacement. Here SIM refers to the simulations and SNN refers to the styled neural network. The sum is over the $N$ particles in the cropped subbox. We also construct the Eulerian density fields, $\delta_{i,\mathrm{SNN}}$ and $\delta_{i,\mathrm{SIM}}$, by distributing the particles to a 3D grid using the trilinear, cloud-in-cell scheme. The second term in the loss is the MSE of the Eulerian densities,
\begin{align}
    L_{\delta} = \frac{1}{N}\sum_{i=1}^{N} |\delta_{i,\mathrm{SNN}} - \delta_{i,\mathrm{SIM}}|^2 \, .
\end{align}
Here the sum is over grid sites, but the number of these is equal to the number of particles in the batch for our choice of Eulerian grid resolution. For the third loss term, we use Eq.~\ref{eq:vel} and compute the MSE for the particle velocities
\begin{align}
    L_{\dot{\vec{\Psi}}} = \frac{1}{N}\sum_{i=1}^{N} |\dot{\vec{\Psi}}_{i,\mathrm{SNN}} - \dot{\vec{\Psi}}_{i,\mathrm{SIM}}|^2 \, .
\end{align}
The final term corresponds to the MSE loss of the Eulerian momentum field. We distribute the particle velocities to an Eulerian vector field mesh with the same resolution as the initial Lagrangian grid and compute,
\begin{align}
    L_{\vec{p}} = \frac{1}{N}\sum_{i=1}^{N} |\vec{p}_{i,\mathrm{SNN}} - \vec{p}_{i,\mathrm{SIM}}|^2 \, ,
\end{align}
where $\vec{p}$ is the Eulerian momentum per particle mass in the grid cells. The final loss function for a snapshot at redshift $z$ is
\begin{align}
L = \frac{1}{1+z} \log(L_{\vec{\Psi}}^6  L_{\delta}^2 L_{\dot{\vec{\Psi}}}^3 L_{\vec{p}}^3) \, .
\end{align}
The exponents are somewhat arbitrary, but they have the effect of weighting the terms of the loss function differently. We have not optimized these exponents, but this choice achieves good training efficiency. The prefactor in the loss function down-weights high-redshift snapshots, where the N-body mapping is easier for the neural network to learn. During the initial phase of training, where we fix the network parameters from the previous, fixed-redshift mode and train only the style parameters, we omit this factor.

We could also choose an Eulerian grid of a higher resolution in order to improve performance on small scales. However, this tends to result in tradeoffs, degrading large-scale accuracy in favour of the smaller scales that dominate the information content in the fields. It may be possible to implement weighting schemes to tailor the loss function to preserve accuracy on larger scales, but we leave this to future work.

\section{Results}
\label{sec:res}

\subsection{Phase space}
\label{ssec:res:ps}

\begin{table}
	\centering
	\begin{tabular}{|c|c|c|c|c|}
		\hline
		$\Omega_{\rm m}$ & $\Omega_{\rm b}$ & $h$    & $n_{\rm s}$ & $\sigma_8$ \\ \hline
		0.3175           & 0.04900          & 0.6711 & 0.9624      & 0.834      \\
		0.1663           & 0.04783          & 0.6173 & 1.1467      & 0.6461     \\
		0.1289           & 0.06325          & 0.7293 & 1.1537      & 0.9489     \\
		0.4599           & 0.04055          & 0.7287 & 0.8505      & 0.7011     \\
		0.4423           & 0.03533          & 0.8267 & 1.0009      & 0.9151     \\ \hline \hline
	\end{tabular}
	\caption{This table lists the cosmological parameters of the five simulations used to test our emulator. The first row corresponds to the Quijote fiducial cosmology, while the other four correspond to extreme corners of the ($\Omega_{\rm m}$, $\sigma_8$) subspace of the 5-parameter Qujoite Latin hypercube.}
	\label{tab:params}
\end{table}

    We evaluate the accuracy of the emulator by constructing the Eulerian density and momentum auto power spectra and cross power spectra for the emulator output and the N-body simulation ground truth. For the density fields $\delta_{\rm SNN}(\vec{x})$ and $\delta_{\rm SIM}(\vec{x})$ we distribute the particles to a $512^3$ mesh using the CIC interpolation scheme and estimate the power spectra,
    \begin{align}
        \langle \delta_{I}(\vec{k}) \delta_{J}(\vec{k}') \rangle = (2\pi)^3 \delta_{\rm D}^{(3)}(\vec{k}+\vec{k}') P_{\delta,IJ}(k) \, ,
    \end{align}
    where $I$ and $J$ can be either SNN or SIM.

    For the momentum fields $\vec{p}_{\rm SNN}(\vec{x})$ and $\vec{p}_{\rm SIM}(\vec{x})$, we distribute the particles weighted by their velocity components to three grids (one for each velocity component) using the CIC scheme and estimator the power spectra
    \begin{align}
        \langle \vec{p}_{I}(\vec{k}) \cdot \vec{p}_{J}(\vec{k}') \rangle = (2\pi)^3 \delta_{\rm D}^{(3)}(\vec{k}+\vec{k}') P_{p,IJ}(k) \, .
    \end{align}
    Note that this involves the dot product between the modes of the two vector fields.

\begin{figure*}
	\includegraphics[width=1\linewidth]{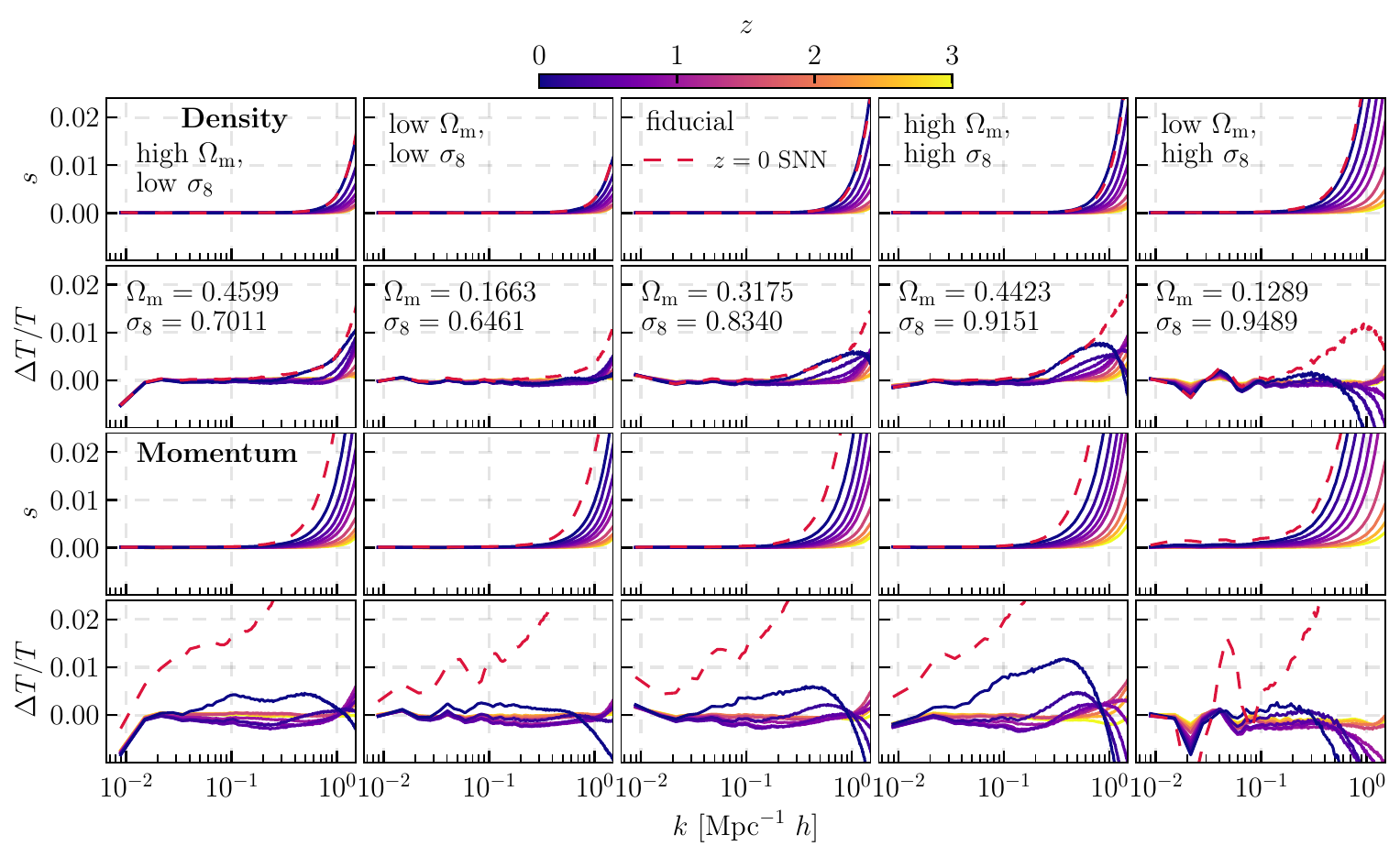}
	\caption{This figure shows the SNN emulator's power spectrum errors. The top row shows the cross-correlation error, or stochasticities, for the density field. The second row shows the amplitude errors, or relative transfer function errors, for the densities. The third and fourth rows show the same for the momentum field. The colour of each curve indicates its redshift, according to the colour bar at the top. The columns show the errors for specific cosmologies, corresponding to extreme corners of the Latin hypercube, and the fiducial cosmology at the center of the hypercube. The red dashed line shows the results from the original, redshift-independent model \citep{Jamieson:2022lqc}, which could only map to $z=0$ and had two independent neural networks for the displacements and velocities.}
	\label{fig:xpks}
\end{figure*}

    We define the stochasticities,
    \begin{align}
    	s = 1 - \frac{(P_{\mathrm{SNN}\times\mathrm{SIM}})^2}{P_{\mathrm{SNN}} P_{\mathrm{SIM}}} \, ,
    \end{align}
    where $P_{\mathrm{SNN}}$ and $P_{\mathrm{SIM}}$ are the auto power spectra for the emulator and the simulation respectively and $P_{\mathrm{SNN}\times\mathrm{SIM}}$ is their cross power spectrum. We also define the fractional transfer function error,
    \begin{align}
    	\frac{\Delta T}{T} = \sqrt{\frac{P_{\mathrm{SNN}}}{P_{\mathrm{SIM}}}} - 1 \, .
    \end{align}
    For individual modes, the stochasticities quantify the error in phases and the transfer function errors quantify the error in the mode amplitudes. Binning the power spectrum, however, mixes contributions from phase and amplitude errors.

    The emulator never encounters any of the 5 test simulations during training. The cosmological parameters for these simulations are listed in Table~\ref{tab:params}. We reran these simulations and output a denser sample of redshifts, $z\in\{0.0, 0.25, 0.5, 0.75, 1.0, 1.5, 2.0, 2.5, 3.0\}$ to test the model performance when interpolating between the five fixed redshifts in the training data.

    In Fig.~\ref{fig:xpks} we show the errors as a function of scale with the color of each curve indicating redshift. The Eulerian density errors originate solely from errors in the particle displaces. Results from the original model from \citet{Jamieson:2022lqc} are shown as the dashed red curves. At redshift $z=0$, the stochasticities of the new time-dependent model are comparable to those of the original model, and the transfer function errors are generally improved over the original model.

    The momentum errors arise from a combination of the particle velocities and displacements, which determine how they are distributed onto the Eulerian grid. The redshift-dependent model achieves significantly reduced errors compared with the original model at $z=0$. This demonstrates the efficacy and constraining power of imposing the physical connection between displacement time dependence and velocity and training directly on the full dynamical N-body phase space.

    The emulator errors depend systematically on cosmology as well as redshift. These dependencies have the same underlying cause: the level of nonlinear clustering. The model performs worse at low redshift because the degree of nonlinear clustering is higher at late times. The model also performs worst on cosmologies with high $\sigma_8$ and low $\Omega_{\mathrm{m}}$, as was found in \citet{Jamieson:2022lqc}. These cosmologies have the highest degree of nonlinearity, and correspondingly, the highest abundances of halos. More specifically, when $\Omega_{\rm m}$ is high, the error has only weak dependence on $\sigma_8$, as seen from the first and fourth columns in Fig.~\ref{fig:xpks}. Conversely, the second and fifth columns of Fig.~\ref{fig:xpks} demonstrate that when $\Omega_{\rm m}$ is low the error depends strongly on $\sigma_8$. Similarly, the first and second columns of Fig.~\ref{fig:xpks} show that when $\sigma_8$ is low the error depends very weakly on $\Omega_{\rm m}$. The two left-most columns of Fig.~\ref{fig:xpks} show that high values of $\sigma_8$ lead to very strong dependence of the error on $\Omega_{\rm m}$.

    The error depends monotonically and correlates positively with $\sigma_8$. The amplitude of small-scale power, and thus the formation of halos is controlled by $\sigma_8$. The highly chaotic virialized motion inside of halos makes it the most difficult regime of the N-body map for the emulator to learn, so increasing $\sigma_8$ tends to degrade the emulator's accuracy. The small-scale errors scale as $k^2$. Since we are dealing with displacements mass is automatically conserved so there can be no $k^{0}$ contribution to the error. The next leading term should scale as $k^2$ if the network predictions are isotropic, so this error scaling confirms that the data augmentation used to enforce isotropy is effective.

    Ref.~\citet{Ibanez:2023vxb} found that parameters used to model small-scale stochasticity and redshift space distortions can absorb much of the excess error for high $\sigma_8$ cosmologies. This significantly mitigates the systematic cosmology-dependence of the error. Since there is a large degree of uncertainty over the values of these parameters, and they will be treated as nuisance parameters and marginalized over, the $\sigma_8$ dependence of the errors may not be a significant issue for inference. This will be tested more rigorously in future work.

\begin{figure*}
    \includegraphics[width=0.495\linewidth]{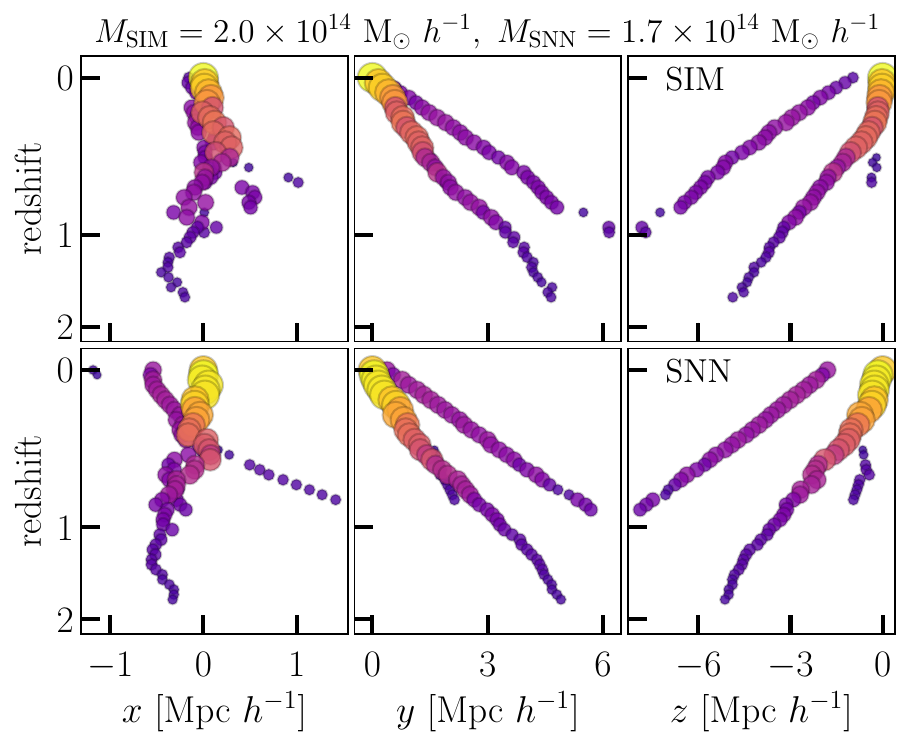}
    \includegraphics[width=0.495\linewidth]{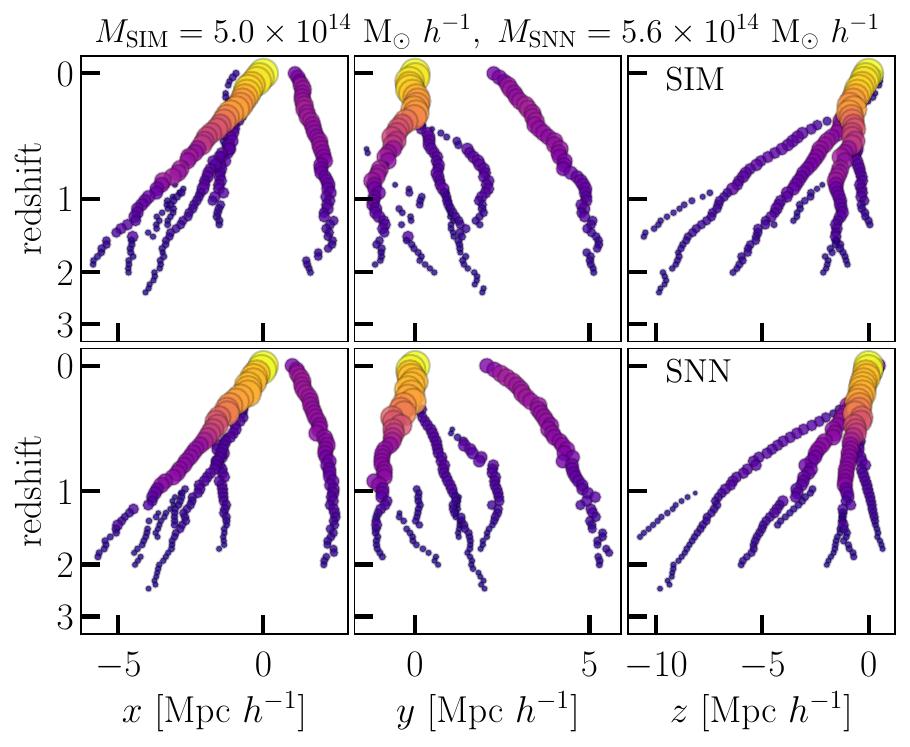}
    \includegraphics[width=0.495\linewidth]{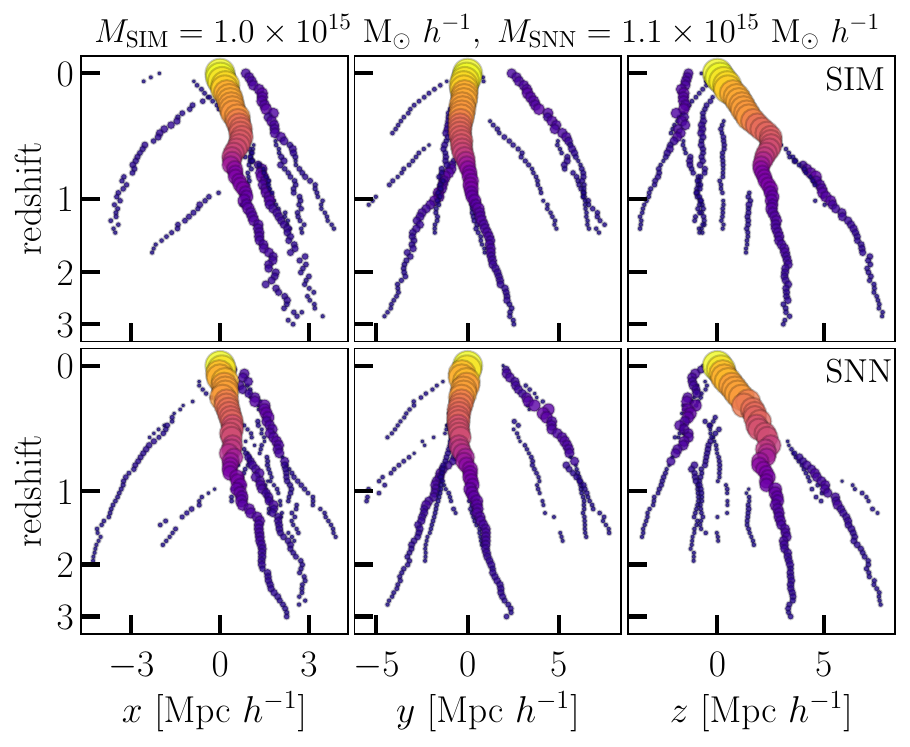}
    \includegraphics[width=0.495\linewidth]{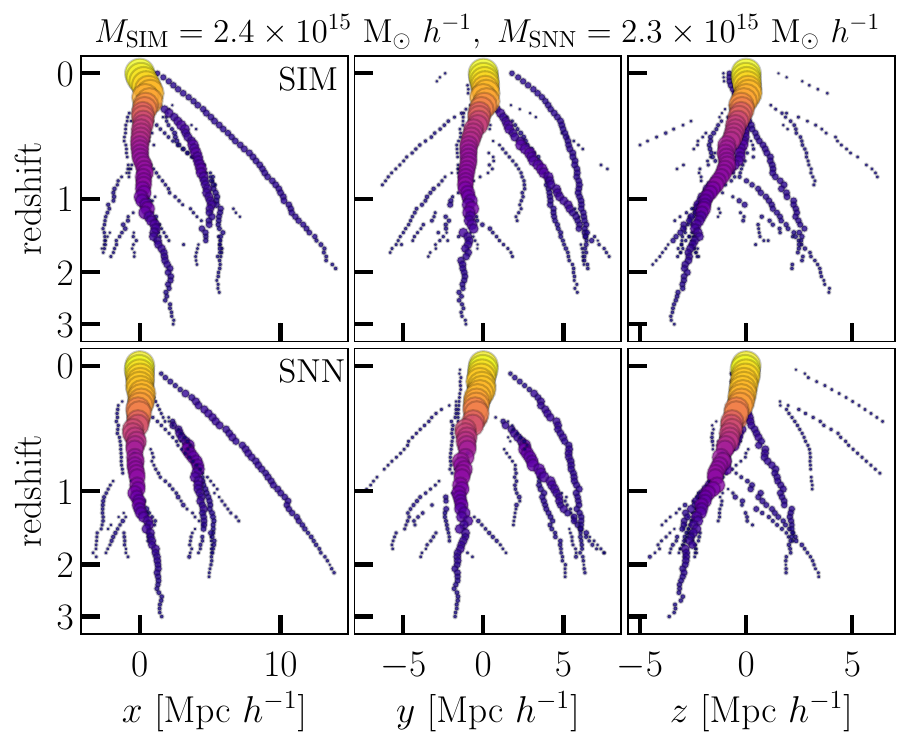}
    \caption{Merger trees for four halos of different masses. Each of the four sets of plots shows the positions of halos along the $x$, $y$, and $z$ directions on the horizontal axes as a function of redshift on the vertical axis. The upper panels show the merger trees from the N-body simulations and the lower panels show the merger trees from the SNN emulator. The main branch and satellites are included. The colour and size of each circle indicate the mass of the object relative to the final halo at redshift zero. The size and color of the points indicate the masses of the halos, but the absolute size is arbitrary, which is why the minor progenitors do not overlap with the major ones at redshift $z=0$.}
    \label{fig:mts}
\end{figure*}

    Another notable feature of the errors is the large-scale bias appearing in the first column of Fig~\ref{fig:xpks}. Here, the transfer function for the most extreme, high $\Omega_{\mathrm{m}}$, low $\sigma_8$ testing simulation has a systematically low transfer function in the first power spectrum bin. Since this cosmology is near the corner of the Latin hypercube where performance is expected to be worst, and since there is a low number of nearby samples in the training dataset here, this indicates the model's limitations when extrapolating in cosmological parameter space. This large-scale bias is absent for cosmologies further away from the edge of the Latin hypercube. In the right-most column of Fig.~\ref{fig:xpks}, we also see oscillatory errors due to the emulator not perfectly predicting the BAO amplitude. These features of the emulator errors are subpercent level, and would likely improve with more training data.

    Errors in higher-order statistics of the original model were explored in Ref/~\cite{Jamieson:2022lqc}. Field-level error diagnostics were also considered in Ref.~\cite{Ibanez:2023vxb}. The redshift-dependent model performs comparably to these for real space density statistics at $z=0$, and significantly better for both redshift space and real-space statistics at higher redshifts, so we do not present these higher-order statistics errors here.

    The errors increase smoothly and monotonically with decreasing redshift. This demonstrates that the emulator interpolates effectively between the small number of fixed-redshift snapshots in its training data.  The model is not overfitting to match the training redshifts, otherwise we would see oscillatory features for errors at intermediate times. A reason that could explain the lack of overfitting for such a small sample size of redshifts is that redshift is not strongly connected to gravitational clustering as a time coordinate. Cosmic time, for example, varies with cosmology at fixed redshift. The network is exposed to a wide sample of cosmic times, due to the range of cosmological parameters in its training data. The network can infer from this data how to accurately interpolate between redshifts at fixed cosmology.

\subsection{Merger Trees}
\label{ssec:res:mt}

A more stringent test of the field-level emulator's accuracy is to analyze the sequences of structure formation that it predicts and check whether this matches the N-body simulations. Structure formation occurs in a bottom-up sequence, where small objects collapse first and then merge to form larger objects, which continue merging and forming increasingly massive objects.

To test our emulator's predictions for formation histories, we reran the fiducial cosmology simulation, outputting 64 snapshots between redshifts $z=3$ and $z=0$. We then ran the Gadget4 \citep{Springel:2020plp} halo finder, subhalo finder {\sc subfind} \citep{Springel:2000qu}, and merger-tree construction algorithm \citep{Springel:2005nw}. Fig. \ref{fig:mts} displays four merger trees for final halos occurring in the emulator prediction and the N-body simulation. In each panel, the vertical axis is redshift, and the three horizontal axes are the $x$, $y$, and $z$ Cartesian coordinates with the final halo at the origin. The N-body results are in the top three panels of each subplot, and the emulator results appear in the bottom three panels. Overall, the merger sequences from the emulator and simulation qualitatively agree. The halos have comparable environments, satellites, and substructures throughout their formation histories.

\begin{figure*}
    \includegraphics[width=1.\linewidth]{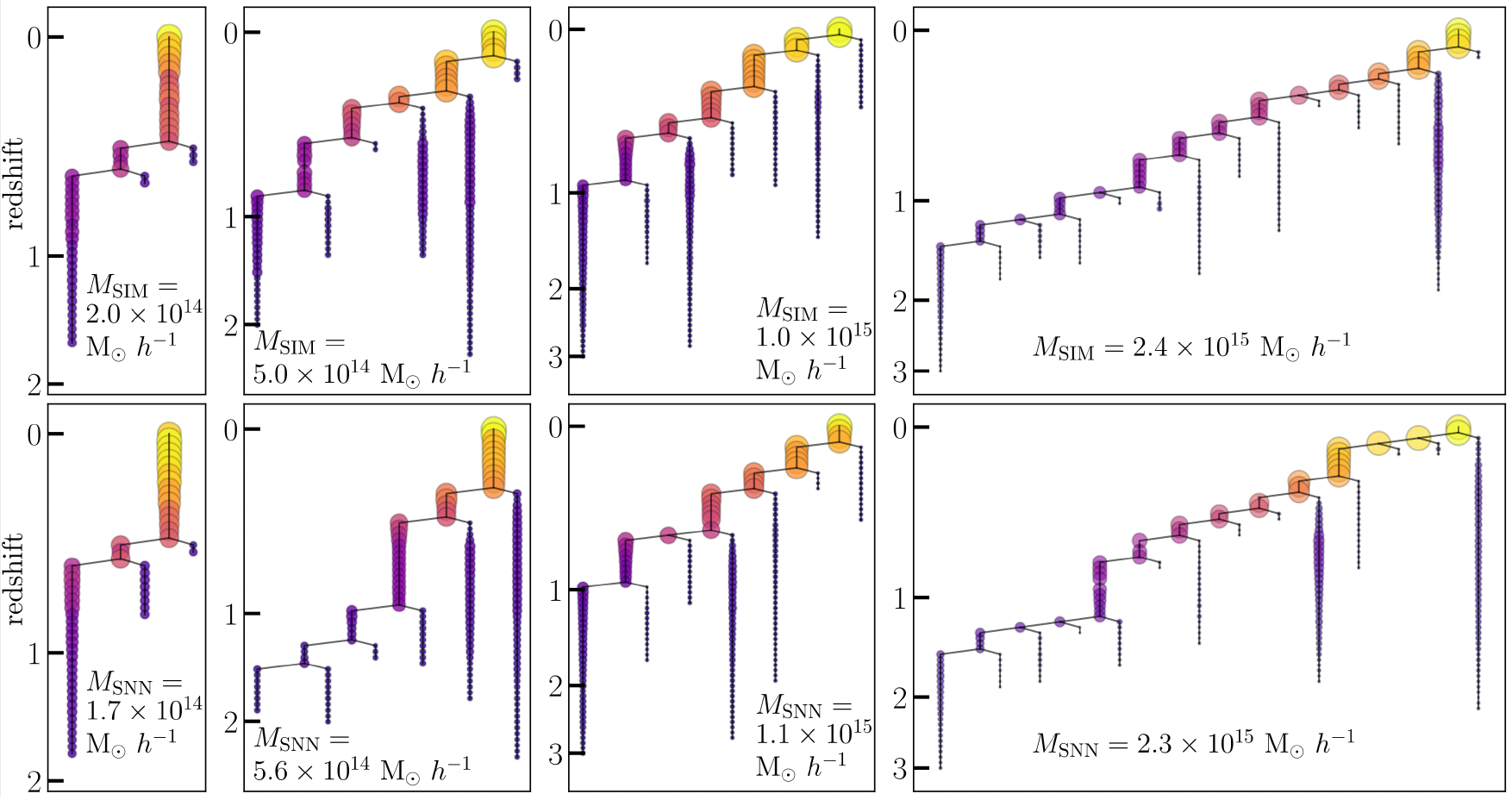}
    \includegraphics[width=0.495\linewidth]{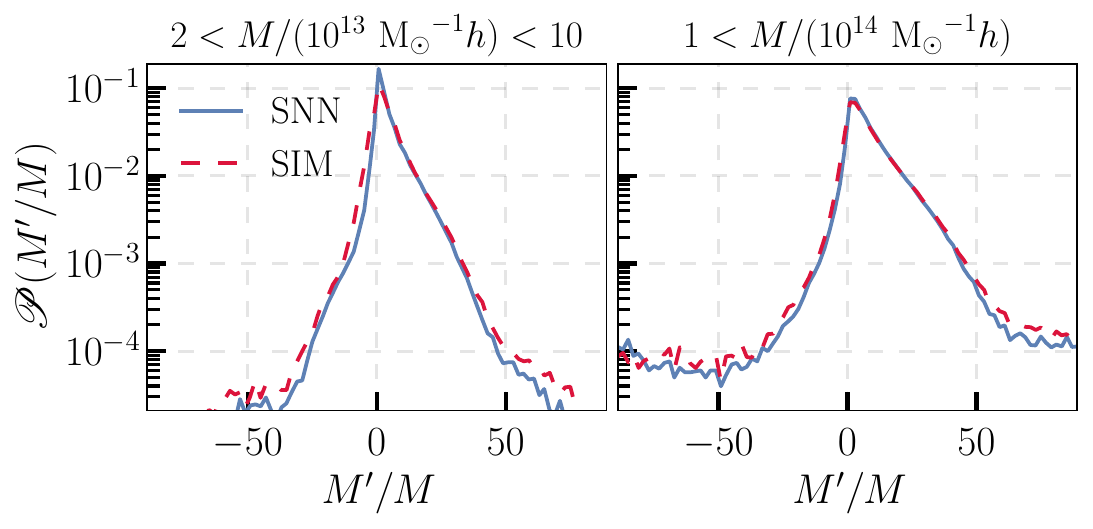}
    \includegraphics[width=0.495\linewidth]{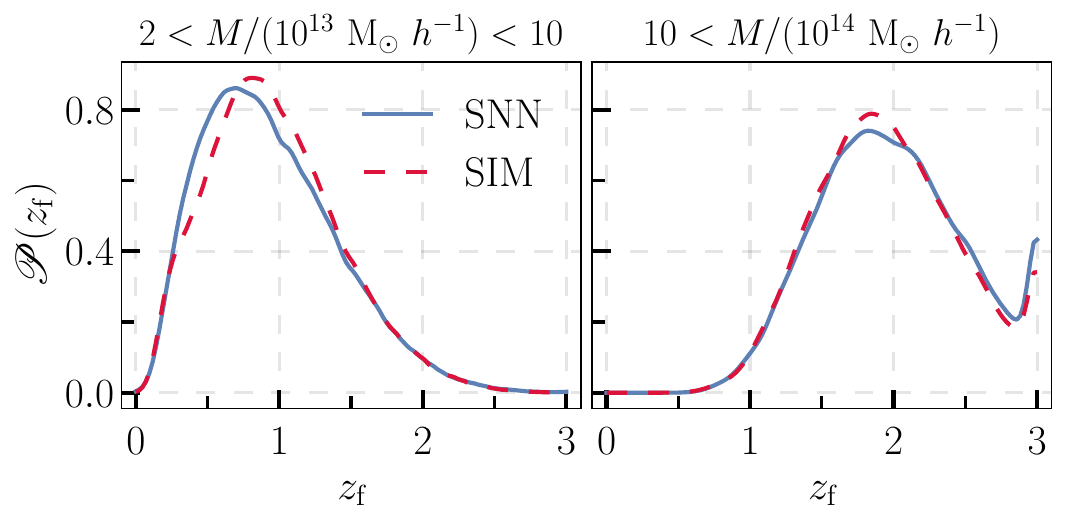}
    \caption{The top two rows show the merger trees for four halos of different masses. The top row shows the simulations merger trees and the second row shows the SNN emulator merger trees for the same halos. The colour and size of each circle indicate the halo's mass relative to the final halo at redshift zero. The bottom row shows the probability densities for the logarithmic rate of mass accretion (left) and the earliest redshift along the main branch (right), for two different halo mass bins.}
    \label{fig:mtstats}
\end{figure*}

A closer inspection of Fig.~\ref{fig:mts} uncovers some differences. The emulator will occasionally merge a satellite that persists at late times in the simulation or miss the merger of a satellite. In the bottom left, the emulator misses a late-time satellite not resolved by the halo finder. In the bottom right, the emulator has merged a satellite into the main branch halo that did not merge in the simulation. In the top left, the emulator has a large satellite with a trajectory in the $x$-direction that does not follow the simulation. The emulator does not perfectly predict all small-scale structure formation details, but it faithfully reproduces the most salient features of the merger histories on $\sim{\rm Mpc}$ scales. A schematic version of these merger trees that more clearly demonstrates the sequences of events is depicted in Fig.~\ref{fig:mtstats}.

Following \citet{Zhang:2023lqi}, we also test the statistical properties among the whole set of merger trees. Specifically, we quantify the statistics of mass accretion rates and progenitor formation times for main branch halos in each merger tree (see \citet{Srisawat:2013vca} for a detailed discussion of merger tree statistics, including those we use here).  Note that a comparison to the results of \citet{Zhang:2023lqi} is not possible since their resolution is significantly higher, making the substructures they resolve much lower in mass. Also, their model is generative, whereas ours is deterministic.

Between two sequential snapshots, the main branch halo in the merger tree will undergo a mass change from $M_i$ in the earlier snapshot at scale factor $a_i$, to $M_{i+1}$ in the later snapshot at scale factor $a_{i+1}$. We define the fractional rate of mass change between these snapshots,
\begin{align}
    \frac{M'}{M} \equiv \frac{\log(M_{i+1} / M_i)}{\log(a_{i+1} / a_i)} \, .
\end{align}
We then estimate the probability density function (PDF) among all main-branch halos in each merger tree between every sequential pair of snapshots. These are displayed in the bottom-left of Fig.~\ref{fig:mtstats} for two different mass bins. The emulator results agree well with the simulation results in the high-mass bin. The plot is on a log scale, so the emulator PDF for the low-mass bin is more sharply peaked and has smaller dispersion compared with the simulation. This discrepancy is unsurprising. The emulator makes errors on $\sim{\rm\!\! Mpc}$ scales similar to the size of a halo. Thus, the emulator predicts fuzzier halos than the simulations, with a more diffuse density profile. This tends to smooth out the mass accretion, leading to more incremental accretion events and fewer major mergers with dramatic changes in mass. This effect is much more pronounced for low-mass halos.

We also define the earliest redshift that a main branch halo appears as $z_f$. The lower-right panels of Fig.~\ref{fig:mtstats} display the PDFs for $z_f$ in two mass bins, which agree between the emulator and the simulations. Since higher-mass halos form from a sequence of merger events of smaller-mass halos, they typically have earlier $z_f$ compared with lower-mass halos. This is reflected in both the simulation and the emulator. For the low-mass bin at around redshift $z=1$, the simulation halos appear slightly earlier compared with those in the emulator, which can also be understood in terms of the emulator's more diffuse halos. This diffuseness means that the halo finder resolves these late-time, low-mass emulator objects at slightly later times than in the simulation. The peak near redshift $z=3$ for the high-mass bin the the bottom-right panel of Fig.~\ref{fig:mtstats} is due to the fact that we have no snapshot earlier than $z=3$, so all halos that form at earlier redshifts are placed in this bin.

These merger tree statistics will be sensitive to the choice of redshift sampling. However, our sampling choice is already much higher than that of the training data. The consistency between the emulator and simulation merger sequences confirms that the emulator accurately models structure formation histories on $~{\rm Mpc}$ scales. This agreement suggests that the emulator can be used alongside semi-analytic models (SAMs) mapping from the initial conditions to galaxies and incorporating hydrodynamic effects of Baryonic physics (see e.g. Refs \citep{Baugh:2006pf,2015ARA&A..53...51S,Barrera:2022jgo} and references therein). This is outside the scope of the current work, but we plan to investigate this in the future.

\section{Conclusion}
\label{sec:conc}

In this work, we presented results and tests of a new field-level emulator for large-scale structure formation. The emulator is a neural network with style parameters that capture cosmology and redshift dependence. The model maps the ZA Lagrangian displacement field at redshift $z$ to the nonlinear displacement field at the same redshift. The model predicts particle velocities through the JVP, forward-mode automatic differentiation of the predicted displacements with respect to time. We trained the model on the six-dimensional phase space of N-body evolution.

The emulator achieves percent-level accuracy on scales of $k\sim {\rm Mpc}~h^{-1}$ in the amplitudes and phases of the density field and momentum field at redshift $z=0$, as shown in Fig.~\ref{fig:xpks}. The emulator performs significantly better at higher redshifts when the nonlinearity is less severe. Our new time-dependent model demonstrates significant improvement over the previous fixed-redshift model \cite{Jamieson:2022lqc} due to the physical constraint imposed during training, that the velocities correspond to time derivatives of the displacements.

We also tested the model's consistency in reproducing the history of structure formation through dark matter merger trees. We found consistency between the emulator predictions and the simulations in terms of halo environments and merger sequences. The emulator smooths out the mass accretion somewhat compared to the simulations, which is most noticeable for low-mass halos.

Using mixed-precision model evaluations, the emulator can process a box of $128^3$ particles (an input of $224^3$ with padding) in under half a second on GPU. The model is parallelized for multi-GPU processing and scales well with box size. The accuracy and speed of our model make it particularly well-suited to field-level inference. The model can also efficiently generate large, high-quality mock datasets for simulations-based inference schemes and high-order N-point statistics analyses. Our field-level emulator provides a power tool for achieving robust and optimal constraints from next-generation cosmological survey data.

\section*{Acknowledgements}
We thank Raul Angulo, Ludvig Doeser, Marcos Pellejero Ibáñez, Jens Jasche, Eiichiro Komatsu, and Volker Springel for their helpful discussions. We thank the Flatiron Scientific Computing Core for providing the computational resources used in this work. The Flatiron Institute is supported by the Simons Foundation.

\bibliographystyle{apsrev4-2}
\bibliography{references}

\end{document}